%% file: Arxiv.tex
\documentclass{article}
\usepackage[utf8]{inputenc}
\usepackage[utf8]{inputenc}
\usepackage[T1]{fontenc}
\usepackage[francais]{babel}
\usepackage{graphicx}
\usepackage{eurosym}
\usepackage{geometry}
\usepackage{mathtools}
\usepackage{float}
\usepackage{multirow}
\usepackage{amsmath,amssymb}
\usepackage{a4wide}
\usepackage[bookmarksopen,bookmarksdepth=2,breaklinks=true]{hyperref}
\usepackage{placeins}
\usepackage{geometry}
\usepackage{eso-pic}
\usepackage{bbold}
\usepackage{natbib}

     \title{Joint model for interval-censored semi-competing events and longitudinal data with subject-specific within and between visits variabilities}
     \author
{Léonie Courcoul$^{1*}$, Catherine Helmer$^1$, 
Antoine Barbieri$^1$  and Hélène Jacqmin-Gadda$^1$
}
\date{}
\begin{document}
\maketitle
\begin{center}
\noindent$^1$Univ. Bordeaux, INSERM, Bordeaux Population Health, U1219, France\\
\end{center}

\begin{abstract}

Dementia currently affects about 50 million people worldwide, and this number is rising. Since there is still no cure, the primary focus remains on preventing modifiable risk factors such as cardiovascular factors. It is now recognized that high blood pressure is a risk factor for dementia. An increasing number of studies suggest that blood pressure variability may also be a risk factor for dementia. However, these studies have significant methodological weaknesses and fail to distinguish between long-term and short-term variability. The aim of this work was to propose a new joint model that combines a mixed-effects model, which handles the residual variance distinguishing inter-visit variability from intra-visit variability, and an illness-death model that allows for interval censoring and semi-competing risks. A subject-specific random effect is included in the model for both variances. Risks can simultaneously depend on the current value and slope of the marker, as well as on each of the two components of the residual variance. The model estimation is performed by maximizing the likelihood function using the Marquardt-Levenberg algorithm. A simulation study validates the estimation procedure, which is implemented in an R package. The model was estimated using data from the Three-City (3C) cohort to study the impact of intra- and inter-visits blood pressure variability on the risk of dementia and death.

\end{abstract}
     
\textbf{Keywords: }Blood pressure, Dementia, Heteroscedasticity, Illness-death model, Interval-censoring, Joint model.

\maketitle
\section{Introduction}
\label{s:intro}

Dementia affects over 50 million people worldwide, and the number of cases continues to rise with increasing life expectancy. With very few treatment's options, prevention of modifiable vascular risk factors remains a crucial action. Numerous studies have already highlighted the link between hypertension and the risk of dementia. Furthermore, an increasing number of studies are focusing on blood pressure variability as a risk factor, independently of blood pressure level \citep{alperovitch_2014, ma_2019, ma_2020}. These studies predominantly examine long-term blood pressure variability, calculated from repeated measures over several years. In comparison, few studies have investigated short or medium-term blood pressure variability \citep{ma_2020}. Moreover, most studies fail to rigorously account for these different types of variability and suffer from methodological weaknesses \citep{decourson_2021}. Typically, variability is calculated as the standard deviation of blood pressure measurements and included as a time-dependent risk factor in a Cox model. This approach introduces bias by neglecting measurement error in the standard deviation of blood pressure and requiring imputation of the standard deviation for all event times. Additionally, blood pressure and its standard deviation are endogenous variables for which the Cox model is not suitable \citep{prentice_1982}.\\
To address these biases, a joint model combining a location-scale linear mixed model with a subject-specific residual variance and a proportional hazards model has been proposed \citep{Gao_2011, Barrett_2019, Courcoul_2024}.  However, this model only accounts for a single blood pressure measurement at each measurement time, whereas in most studies, blood pressure is measured at least twice at each visit. From a clinical perspective, it would be interesting to assess whether short-term (intra-visit) variability predicts dementia, as its measurement is straightforward.\\
Furthermore, given the common risk factors between dementia and death, it is important to consider the competing risk of death in such studies. However, in cohort studies, the onset age of dementia is not precisely known since dementia is only diagnosed at scheduled visit times. This introduces uncertainty about the exact time of dementia onset, which occurs between the last visit where the individual was seen without symptoms and the visit of diagnosis. Moreover, an individual may develop dementia and die between two visits, and thus dementia may not have been diagnosed. Consequently, when interval censoring is not rigorously accounted for, the risk of dementia may be underestimated \citep{leffondre_2013}. To address this issue, \cite{joly_2002} proposed an illness-death model to adjust both the risk of dementia and death, taking into account interval censoring. Later, \cite{rouanet_2016} extended this work to consider longitudinal data via a joint latent class model. However, this latter work does not allow to consider individual-specific variability and to assess its impact on event risk.\\
The aim of this work is to propose a joint model combining a location-scale mixed model decomposing individual short- and long-term residual variance and an illness-death model dealing with both interval censoring and left truncation. In this model, the risks of each event can depend simultaneously on the current value, the slope and both variabilities of the marker.\\
The paper is organized as follows. Section 2 describes the model and the estimation procedure that is then assessed in a simulation study in Section 3. In section 4, the model is applied to the data from the Three-City (3C) cohort \citep{3c_2003} to study the impact of within and between visits blood pressure variabilities on the risk of dementia and death. Finally, Section 5 concludes this work with some elements of discussion.

\section{Method}
Let us consider a sample of $N$ individuals. For each subject $i$ $(i = 1,\ldots,N)$, $Y_{ijl}$ is the marker value for measure $l$ $(l = 1,\ldots,n_{ij})$, at visit $j$ $(j = 1,\ldots,n_i)$ and time $t_{ij}$. For each visit $j$, subject $i$ can have $n_{ij}$ measurements of the longitudinal marker. $Y_i$ is a vector of dimension $\sum_{j=1} ^{n_{i}} n_{ij}$ containing all marker measurements for the subject $i$.\\
We denote $T_i^{Dem}$ the unobserved age at dementia onset and $T_i^{Death}$ the age of death. We assume that age at dementia onset is interval-censored while age at death is known since in most cohorts the exact age at death can be collected. The vector of collected data for time-to-events is given by $D_i = (T_{0i}, L_i, R_i, \delta_i^{Dem}, T_i, \delta_i^{Death})^\top$ where $T_{0i}$ is the age at inclusion, $L_i$ is the age at the last visit where the subject was seen free of dementia, $R_i$ is the age at the visit of diagnosis if the subject was diagnosed with dementia (in case of no diagnosis of dementia, $R_i$ is undefined), $T_i$ is the minimum between the age at death and the age of right censoring (e.g. the end of the follow-up), $\delta_i^{Dem}$ is the indicator of dementia diagnosis and $\delta_i^{Death}$ is the indicator of death. 

\subsection{Joint model with inter and intra-visit individual variabilities}
We propose a joint modelling for a longitudinal outcome and semi-competing events using a shared random-effect approach. The longitudinal submodel is defined by a location-scale linear mixed-effect model decomposing within and between individual residual variance:
\begin{equation}
\left\{
    \begin{array}{ll}
         Y_{ijl} =  \widetilde{Y}_i(t_{ij}) + \epsilon_{ij} + \nu_{ijl} = X_{ij}^{\top} \beta+Z_{ij}^{\top} b_{i}+\epsilon_{ij} + \nu_{ijl}, \\
        \epsilon_{ij} \sim \mathcal{N}(0,\sigma_i^2) \hspace{3mm} \text{with} \hspace{3mm} \log(\sigma_i)  = \mu_\sigma + \tau_{\sigma i},\\
        \nu_{ijl} \sim \mathcal{N}(0,\kappa_i^2) \hspace{3mm} \text{with} \hspace{3mm} \log(\kappa_i)  = \mu_\kappa + \tau_{\kappa i},\\
    \end{array}
\right.  
 \label{Mixed}
\end{equation}
with $X_{ij}$ and $Z_{ij}$ two vectors of explanatory variables for subject $i$ at visit $j$, respectively associated with the fixed-effect vector $\beta$ and the subject-specific random-effect vector $b_i$, and $\mu_\sigma$ and $\mu_\kappa$ two fixed effects associated with the intercept for the between visits and within visit variabilities respectively. The subject-specific random-effect $b_i$ and $\tau_i = (\tau_{\sigma i},\tau_{\kappa i})^\top$ are both normally distributed and could be supposed to be correlated, such as 

\begin{equation}
   \quad\left(\begin{array}{c}
b_{i} \\
\tau_{i} \\
\end{array}\right) \sim \mathcal{N}\left(\left(\begin{array}{c}
0 \\
0 \\
\end{array}\right),\left(\begin{array}{ccc}
\Sigma_{b} & \Sigma_{\tau b}\\
\Sigma_{\tau b}^{\top} & \Sigma_{\tau}
\end{array}\right)\right) 
\end{equation}

\noindent To be able to account for the interval censoring of dementia, the risks of dementia and death are modeled according to an illness-death model (Figure \ref{FigCI}A ). The transition intensities from state $k \in \{0,1\}$ to state $l \in \{1,2\}$ are defined by a proportional hazards model under the Markovian hypothesis:
\begin{equation}
    \lambda_{i}^{kl}(t|b_i, \tau_i)=\lambda_{0}^{kl}(t) \exp \left(W_{i}^{kl^\top} \gamma^{kl}+\alpha_{1}^{kl}\tilde{y}_i(t)+
     \alpha_{2}^{kl}\tilde{y}'_i(t)+ \alpha_{\sigma }^{kl} \sigma_i +  \alpha_{\kappa }^{kl} \kappa_i \right)
    \label{Surv}
\end{equation}
with $\lambda_{0}^{kl}(t)$ the baseline risk function, $W_{i}^{kl}$ a vector of baseline covariates associated with the regression coefficient $\gamma^{kl}$, and $\alpha_{1}^{kl}$, $\alpha_{2}^{kl}$, $\alpha_{\sigma}^{kl}$, and $\alpha_{\kappa}^{kl}$ the regression coefficients associated with the current value $\tilde{y}_i(t)$, the current slope $ \frac{\partial \widetilde{y_i}(t)}{\partial t}=\tilde{y}'_i(t)$, the long-term residual variability $\sigma_i$ and the short-term residual variability $\kappa_i$, respectively. Different parametric forms for the baseline risk functions can be considered, such as exponential, Weibull, or for more flexibility, a B-splines base.

\subsection{Estimation Procedure}
\subsubsection*{Log-likelihood\\}

\begin{figure}
    \centering
    \includegraphics{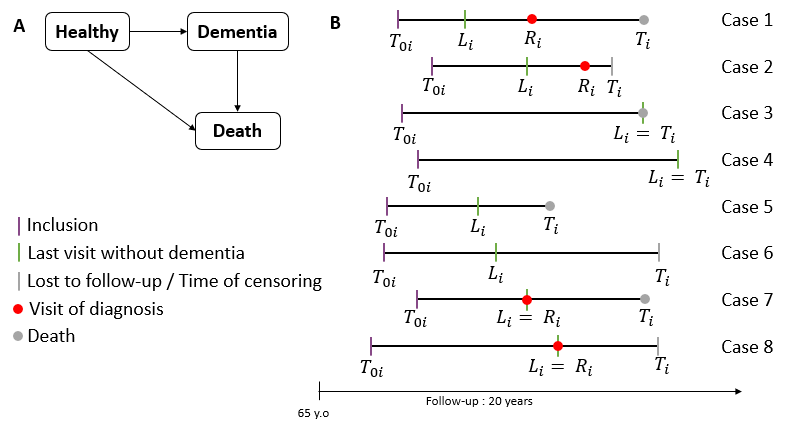}
    \caption{A. Illness-death model. B. Possible patterns of dementia and death.}
    \label{FigCI}
\end{figure}

Let $\theta$ be the set of parameters to be estimated including parameters of the Cholesky decomposition of the covariance matrix of the random effects, $\beta$, $\mu = (\mu_\sigma, \mu_\kappa)^\top$, $\alpha = (\alpha_{1}^{kl},\alpha_{2}^{kl},\alpha_{\sigma}^{kl}, \alpha_{\kappa}^{kl})^\top$, $\gamma = (\gamma^{kl})^\top = (\gamma_{01}, \gamma_{02}, \gamma_{12})^\top$ for $k\in\{0,1\}$ and $l \in \{1,2\}$ and the parameters of the three baseline risk functions. Considering the frequentist approach, the parameter estimation is obtained by maximizing the likelihood function. Under the assumption of independence between $D_i$ and $Y_i$ conditionally on random effects and assuming independent censoring, the contribution of individual $i$ to the marginal likelihood is defined by:
\begin{eqnarray}
\mathcal{L}_i(\theta;Y_i,D_i) & = & \int f(Y_i|b_i,\tau_i;\theta)f(D_i|b_i,\tau_i;\theta)f(b_i,\tau_i;\theta)db_id\tau_i,
\label{log_ll}
\end{eqnarray}
and in case of delayed entry, the log-likelihood is divided by the probability of being alive and healthy at entry:

\begin{equation*}
    \mathcal{L}_i^{DE}(\theta;Y_i,D_i) = \frac{\mathcal{L}_i(\theta;Y_i,D_i)}{\int \exp(-\Lambda_{01i}(T_{0i}|b_i,\tau_i;\theta)-\Lambda_{02i}(T_{0i}|b_i,\tau_i;\theta))f(b_i,\tau_i;\theta)db_id\tau_i}
\end{equation*}

with:
\begin{itemize}
    \item $f(b_i,\tau_i;\theta)$ is a multivariate Gaussian density
    \item $f(Y_i|b_i,\tau_i;\theta)=\prod_{j=1}^{n_i}f(Y_{ij}|b_i,\tau_{\sigma i},\tau_{\kappa i};\theta)$ where $f(Y_{ij}|b_i,\tau_i;\theta)$ is a multivariate Gaussian density with symmetric covariance matrix $\Sigma = \sigma_i^2\mathbb{1}_{n_i}^\top\mathbb{1}_{n_i} + \kappa_i^2 I_{n_i}$, with $\mathbb{1}_{n_i}$ the unity vector of size $n_i$ and $I_{n_i}$ the identity matrice of size $n_i\times n_i$.
    \item$\Lambda_{i}^{01}(T_{0i}|b_i,\tau_i;\theta)$ and $\Lambda_{i}^{02}(T_{0i}|b_i,\tau_i;\theta)$ are the cumulative risk functions respectively for transition Healthy to Dementia and Healthy to Death with $\Lambda_{i}^{0k}(T_{0i}|b_i,\tau_i;\theta) = \int_0^{T_{0i}} \lambda_i^{0k}(t\vert b_i, \tau_i ;\theta) dt$ for $k=1,2$ and $\lambda_i^{0k}(t\vert b_i, \tau_i ;\theta)$ defined in equation \eqref{Surv}
    \item $f(D_i|b_i,\tau_i;\theta)$ is the survival part of the individual contribution to the likelihood that depends on the subject trajectory as illustrated on Figure \ref{FigCI}B.
\end{itemize} 
To present the different definitions of $f(D_i|b_i,\tau_i;\theta)$ according to subjects observations, we deliberately omit the conditioning on random effects and parameters for ease of notation:\\

\noindent\textit{Cases 1 and 2:} subject healthy and alive until $L_i$, diagnosed with dementia at $R_i$, remained alive until $T_i$, possibly died at $T_i$: 
\begin{eqnarray*}
f(T_{0i}, L_i, R_i, \delta_i^{Dem} = 1, T_i, \delta_i^{Death}) &=& \int_{L_i}^{R_i} e^{-\Lambda_{i}^{01}(u) - \Lambda_{i}^{02}(u)}\lambda_{i}^{01}(u)e^{-(\Lambda_{i}^{12}(T_i)-\Lambda_{i}^{12}(u))}\lambda_{i}^{12}(T_i)^{\delta_i^{Death}}du
\end{eqnarray*}

\noindent\textit{Cases 3 and 4:} subject healthy and alive until $T_i$ ($T_i = L_i$), possibly died at $T_i$: 
\begin{eqnarray*}
f(T_{0i}, L_i, R_i, \delta_i^{Dem} = 0, T_i, \delta_i^{Death}) &=& e^{-\Lambda_{i}^{01}(T_i) - \Lambda_{i}^{02}(T_i)}\lambda_{i}^{02}(T_i)^{\delta_i^{Death}}
\end{eqnarray*}

\noindent\textit{Cases 5 and 6:} subject healthy at his/her last visit $L_i = R_i < T_i$, remained alive until $T_i$ and possibly died at $T_i$: 
\begin{eqnarray*}
f(T_{0i}, L_i, R_i, \delta_i^{Dem} = 0, T_i, \delta_i^{Death}) &=& e^{-\Lambda_{i}^{01}(T_i) - \Lambda_{i}^{02}(T_i)}\lambda_{i}^{02}(T_i)^{\delta_i^{Death}} +\\
&&\int_{L_i}^{T_i} e^{-\Lambda_{i}^{01}(u) - \Lambda_{i}^{02}(u)}\lambda_{i}^{01}(u)e^{-(\Lambda_{i}^{12}(T_i)-\Lambda_{i}^{12}(u))}\lambda_{i}^{12}(T_i)^{\delta_i^{Death}}du
\end{eqnarray*}
The likelihood accounts for the two possible trajectories of this subject: either direct transition from healthy to death or unobserved transition to dementia between $L_i$ and $T_i$.\\

\noindent\textit{Cases 7 and 8:} subject diagnosed with dementia at $L_i = R_i$ (exact date of dementia onset), remained alive until $T_i$, possibly died at $T_i$: 
\begin{eqnarray*}
f(T_{0i}, L_i, R_i, \delta_i^{Dem} = 1, T_i, \delta_i^{Death}) &=& e^{-\Lambda_{i}^{01}(L_i) - \Lambda_{i}^{02}(L_i)}\lambda_{i}^{01}(L_i)e^{-(\Lambda_{i}^{12}(T_i)-\Lambda_{i}^{12}(L_i))}\lambda_{i}^{12}(T_i)^{\delta_i^{Death}}
\end{eqnarray*}
These cases do not occur in the context of dementia but could arise in other types of events.

\subsubsection*{Optimisation\\}

\noindent Since the integral over the random effects does not have an analyticial solution, it is computed using a Quasi Monte Carlo (QMC) approximation \citep{pan_quasi-monte_2007}, employing deterministic quasi-random sequences. This leads to the equation \eqref{log_ll} being approximated by: 
\begin{equation*}
   \mathcal{L}_i^{DE}(\theta;Y_i,D_i)  \simeq \frac{\frac{1}{S} \sum_{s = 1}^Sp(Y_i,D_i|b_i^s,\tau_i^s;\theta)}{\frac{1}{S}\sum_{s = 1}^S \exp(-\Lambda_{i}^{01}(T_{0i}|b_i^s,\tau_i^s;\theta)-\Lambda_{i}^{02}(T_{0i}|b_i^s,\tau_i^s;\theta))} 
\end{equation*} 
where $(b_i^1,...,b_i^S)$ and $(\tau_{\sigma i}^1,...,\tau_{\sigma i}^S,\tau_{\kappa i}^1,...,\tau_{\kappa i}^S)$ are draws of a S-sample in the sobol sequel for the distribution $f(b_i,\tau_{\sigma i}, \tau_{\kappa i};\theta)$. Moreover, to approximate the cumulative risk functions, we use the Gauss-Kronrod quadrature approximation with 15 points \citep{gonnet_2012}.\\

\noindent Parameter estimation is obtained by maximizing the log-likelihood function $\ell(\theta;Y,D) = \log\left( \prod_{i=1}^N 
\mathcal{L}_i(\theta;Y_i,D_i)\right)$. The maximization is performed using the \texttt{marqLevAlg} R-package based on the Marquardt-Levenberg algorithm \citep{philipps_2021}.\\

The variances of the estimates are estimated by the inverse of the Hessian matrix computed by finite differences. The variances of the parameters of the random effects covariance matrix are obtained from those of the parameters of the Cholesky decomposition using the Delta-Method \citep{meyer_2013}.\\

In order to insure precise estimates of parameters and their standard error and to limit computation time, the estimation is performed in three steps:
\begin{enumerate}
    \item \textbf{Parameters initialisation}. estimation of the joint model without handling interval censoring:
    \begin{itemize}
        \item taking the middle of the interval as time of dementia onset and assuming that subjects who died before diagnosis of dementia directly made the transition health to death;
        \item using the estimated parameters of the estimation from the longitudinal model (equation \eqref{Mixed})
    \end{itemize}
    \item \textbf{Parameters estimation} of the joint model accounting for interval censoring with a sufficient number $S1$ of QMC to avoid bias using the estimated parameters from step $(1)$;
    \item \textbf{Precision improvement} with a number $S2 > S1$ of QMC draws using the estimated parameters from step $(2)$.
\end{enumerate}
The choice of the number of QMC draws $S1$ and $S2$ has a notable effect on computational time. Therefore, for model selection, we suggest that users compare different models using the results of step (2) (using likelihood or information criteria) with a small value for $S1$ and perform step (3), which entails a larger number of QMC draws, solely for the final selected model. 

\section{Simulations}
In order to evaluate the performance of the estimation procedure and compare the estimations with those obtained using a joint model that does not account for interval censoring we carried out simulations driven by the application.

\subsection{Design}

For each subject, age at entry is generated from a beta distribution, over a window of 65 to 85 years old. Visit times are generated using a uniform distribution centered around each specified time, with a variation of $\pm$month in either direction. For each visit time, two measurements ($l =1,2$) of the marker are generated, using a location-scale linear mixed-effects model with fixed and random intercept and slope, and two heterogeneous variances:
\begin{equation}
\left\{
    \begin{array}{ll}
         Y_{ijl} =  \widetilde{Y}_i(t_{ij}) + \epsilon_{ij} + \nu_{ijl} = \beta_0 + b_{0i} +(\beta_1+b_{1i})\times t_{ij}+\epsilon_{ij} + \nu_{ijl} \\
        \epsilon_{ij} \sim \mathcal{N}(0,\sigma_i^2) \hspace{3mm} \text{with} \hspace{3mm} \log(\sigma_i)  = \mu_\sigma + \tau_{\sigma i},\\
        \nu_{ijl} \sim \mathcal{N}(0,\kappa_i^2) \hspace{3mm} \text{with} \hspace{3mm} \log(\kappa_i)  = \mu_\kappa + \tau_{\kappa i},\\
    \end{array}
\right.
 \label{MixedSimu}
\end{equation}
where $t$ is a scaled transformation of age $t$ : $t = \frac{age-65}{10}$. 



The follow-up visits, $(t_{ij})_{j=(1,\ldots,n_{i})}$, are scheduled regularly from inclusion until the minimum between death and the administrative right-censoring which is 20 years after inclusion. The process of age of dementia onset and death generation is described on Supplementary Materials with each function of transition defined by the following proportional hazards models:
\begin{equation}
    \lambda_{i}^{kl}(t)=\lambda_{0}^{kl}(t) \exp \left(\alpha_{1}^{kl}\tilde{y}_i(t)+
     \alpha_{2}^{kl}\tilde{y}'_i(t)+ \alpha_{\sigma }^{kl} \sigma_i +  \alpha_{\kappa }^{kl} \kappa_i \right)
    \label{SimuSurv}
\end{equation}
with $\lambda_{0}^{kl}(t) = \eta^{kl} t^{\eta^{kl}-1}e^{\alpha_{0}^{kl}}$ being a Weibull function. 

Three scenarii were performed, varying the parameters and the number of repeated measures:
\begin{itemize}
    \item Scenario A: parameters driven by the estimation on the model defined by \eqref{MixedSimu} and \eqref{SimuSurv} on the 3C cohort, and visit times driven by 3C cohorts: at 2, 4, 7, 10, 12, 14 and 17 years from inclusion.
    \item Scenario B: same parameters as in Scenario A but visit times at 4, 8, 12 and 16 years from inclusion to assess the impact of the interval between visits on the results.
    \item Scenario C: increased signal on interest parameters to simulate a positive effect of the current value and between-visits variability on the risk of dementia, and same visit times as in Scenario A.
\end{itemize}
For each scenario, 500 samples of 1000 subjects have been generated. We also estimated a naive joint illness-death model without handling interval censoring considering the exact time of dementia onset known as the middle of the interval between the last visit without symptoms and the visit of diagnosis for diagnosed individuals and considering that subjects who died before diagnosis of dementia directly made the transition healthy to death.

\subsection{Results}

\begin{table}[]
\caption{Results of the simulation study for Scenario A, comparing estimates of the joint illness-death model for interval-censored events and the naive joint illness-death model. A total of 500 samples of 1000 subjects were generated with a joint illness-death model with visit times driven by 3C design. ASE is the asymptotic standard error, ESE is the empirical standard error and the coverage rate is calculated from the 95\% confidence interval.}
\centering
\label{SA}
\scalebox{0.8}{
\begin{tabular}{lccccccccccc}
\hline
                                                                                                                                                                      
                           & \multicolumn{1}{l}{} & \textbf{}            & \multicolumn{4}{c}{\textbf{Interval censoring$^a$}}                                           & \multicolumn{1}{l}{} & \multicolumn{4}{c}{\textbf{Naive model$^b$}}                                                 \\ \cline{4-7} \cline{9-12} 
Parameter                  & \multicolumn{1}{l}{} & $\theta$             & $\hat{\theta}$       & ESE                  & ASE                  & CR (95\%)            &                      & $\hat{\theta}$       & ESE                  & ASE                  & CR (95\%)            \\ \hline
\multicolumn{2}{l}{\textit{Longitudinal process}} & \multicolumn{1}{l}{} & \multicolumn{1}{l}{} & \multicolumn{1}{l}{} & \multicolumn{1}{l}{} & \multicolumn{1}{l}{} & \multicolumn{1}{l}{} & \multicolumn{1}{l}{} & \multicolumn{1}{l}{} & \multicolumn{1}{l}{} & \multicolumn{1}{l}{} \\ \hline
\textit{Intercept}         & $\beta_0$            & $14.0$                 & $14.00$                & $0.10$                 & $0.10$               & $95.2$                 &                      & $13.98$                & $0.10$                 & $0.10$                 & $94.2               $  \\
\textit{Slope}             & $\beta_1$            & $0.17$                 & $0.170$                & $0.064$                &$ 0.063               $ & $95.6$                 &                      & $0.193               $ & $0.062$                & $0.061$                & $92.8$                 \\
\textit{Variability inter} & $\mu_\sigma$         & $0.30                $ & $0.300$                & $0.017$                & $0.017$                & $94.8$                 &                      & $0.300$                & $0.017$                & $0.017$                & $94.8$                 \\
\textit{Variability intra} & $\mu_\kappa$         & $-0.23               $ &$ -0.231$               & $0.013$                & $0.013$                & $94.0$                 &                      & $-0.232 $              & $0.013 $               & $0.013$                & $94.4  $               \\ \hline
\multicolumn{2}{l}{\textit{Transition 0-1}}       & \multicolumn{1}{l}{} & \multicolumn{1}{l}{} & \multicolumn{1}{l}{} & \multicolumn{1}{l}{} & \multicolumn{1}{l}{} & \multicolumn{1}{l}{} & \multicolumn{1}{l}{} & \multicolumn{1}{l}{} & \multicolumn{1}{l}{} & \multicolumn{1}{l}{} \\ \hline
\textit{Current value}     & $\alpha_1^{01}$      & $-0.06$                & $-0.059$               & $0.045               $ & $0.045$                & $95.0$                 &                      & $-0.071$               & $0.045$                & $0.045$                & $95.0$                 \\
\textit{Current slope}     & $\alpha_2^{01}$      & $0.0                 $ & $-0.002$               & $0.093$                & $0.094 $               & $94.6$                 &                      & $-0.017$               & $0.093$                & $0.093$                & $95.2 $                \\
\textit{Inter variability} & $\alpha_\sigma^{01}$ & $0.50                $ & $0.521$                & $0.320$                & $0.297$                & $94.4$                 &                      & $0.569$                & $0.317$                & $0.300$                & $95.4$                 \\
\textit{Intra variability} & $\alpha_\kappa^{01}$ & $0.01$                 & $-0.013$               & $0.478   $             & $0.453$                & $94.0   $              &                      &$ -0.005 $              & $0.479   $             &$ 0.461 $               & $94.4$                 \\
\textit{Weibull}           & $\sqrt{\eta^{01}}$   & $2.00$                 & $2.012 $               &$ 0.055 $               & $0.057$                & $94.8 $                &                      & $1.907$                & $0.056  $              & $0.058               $ & $60.5$                 \\
                           & $\zeta^{01}$         & $-4.00               $ & $-4.080$               & $0.764$                & $0.781$                & $94.8$                 &                      & $-3.815$               & $0.744$                & $0.776$                & $95.0 $                \\ \hline
\multicolumn{2}{l}{\textit{Transition 0-2}}       & \multicolumn{1}{l}{} & \multicolumn{1}{l}{} & \multicolumn{1}{l}{} & \multicolumn{1}{l}{} & \multicolumn{1}{l}{} & \multicolumn{1}{l}{} & \multicolumn{1}{l}{} & \multicolumn{1}{l}{} & \multicolumn{1}{l}{} & \multicolumn{1}{l}{} \\ \hline
\textit{Current value}     & $\alpha_1^{02}$      & $-0.10               $ & $-0.107$               & $0.051 $               & $0.052  $              & $95.2$                 &                      &$ -0.069  $             & $0.038 $               &$ 0.039               $ & $86.7     $            \\
\textit{Current slope}     & $\alpha_2^{02}$      & $-0.40$                & $-0.411$               & $0.119               $ & $0.115$                & $95.8 $                &                      &$ -0.293 $              & $0.082  $              &$ 0.084               $ & $72.1    $             \\
\textit{Inter variability} & $\alpha_\sigma^{02}$ & $0.46$                 & $0.440$                & $0.368$                & $0.347$                & $95.6$                 &                      &$ 0.412$                & $0.264$                & $0.265$                & $96.0$                 \\
\textit{Intra variability} & $\alpha_\kappa^{02}$ & $0.21$                 & $0.181$                & $0.553$                & $0.516$                & $94.6$                 &                      & $0.133 $               & $0.387$                & $0.394$                & $96.8$                 \\
\textit{Weibull}           & $\sqrt{\eta^{02}}$   & $1.70$                 & $1.708$                & $0.060$                & $0.060$                & $94.6$                 &                      & $1.822$                & $0.051$                & $0.049$                & $32.3$                 \\
                           & $\zeta^{02}$         & $-2.50$                & $-2.398$               & $0.906$                & $0.897$                & $95.0$                 &                      & $-3.031$               & $0.691$                & $0.695$                & $89.4$                 \\ \hline
\multicolumn{2}{l}{\textit{Transition 1-2}}       & \multicolumn{1}{l}{} & \multicolumn{1}{l}{} & \multicolumn{1}{l}{} & \multicolumn{1}{l}{} & \multicolumn{1}{l}{} & \multicolumn{1}{l}{} & \multicolumn{1}{l}{} & \multicolumn{1}{l}{} & \multicolumn{1}{l}{} & \multicolumn{1}{l}{} \\ \hline
\textit{Current value}     & $\alpha_1^{12}$      & $0.04$                 & $0.047$                & $0.056$                & $0.056               $ & $94.8$                 &                      & $0.027$                & $0.046$                & $0.055               $ & $97.6$                 \\
\textit{Current slope}     & $\alpha_2^{12}$      & $0.02$                 & $0.011$                & $0.120$                & $0.120$                & $94.8$                 &                      & $-0.021$               & $0.081 $               & $0.107               $ & $98.6$                 \\
\textit{Inter variability} & $\alpha_\sigma^{12}$ & $-0.12               $ &$ -0.127 $              & $0.352 $               & $0.335$                & $95.0$                 &                      & $0.038$                & $0.211 $               & $0.278               $ & $97.0 $                \\
\textit{Intra variability} & $\alpha_\kappa^{12}$ & $-0.18               $ & $-0.158$               & $0.571$                & $0.531$                & $93.8$                 &                      & $-0.142$               & $0.372$                & $0.463$                & $98.2$                 \\
\textit{Weibull}           & $\sqrt{\eta^{12}}$   & $1.70                $ & $1.717 $               & $0.104$                & $0.106 $               & $94.2 $                &                      & $1.788$                & $0.095 $               &$0.103               $ & $87.2 $                \\
                           & $\zeta^{12}$         & $-2.20               $ & $-2.385$               & $1.022 $               & $1.005  $              & $95.0$                 &                      & $-2.929$               & $0.823 $               & $0.976               $ & $95.2$      
                         \\  \hline
\end{tabular}}\\
\footnotesize{
$^a$496 samples with convergence criteria fulfilled; $^b$499 samples with convergence criteria fulfilled.}
\end{table}



\begin{table}[]
\caption{Results of the simulation study for Scenario B, comparing estimates of the joint illness-death model for interval-censored events and the naive joint illness-death model. A total of 500 samples of 1000 subjects were generated with a joint illness-death model with visit times every 4 years. ASE is the asymptotic standard error, ESE is the empirical standard error and the coverage rate is calculated from the 95\% confidence interval.}
\centering
\label{SB}
\scalebox{0.8}{
\begin{tabular}{lccccccccccc}
\hline
                                                                                                                                                                                                                         
                           & \multicolumn{1}{l}{} & \textbf{}            & \multicolumn{4}{c}{\textbf{Interval censoring$^a$}}                                           & \multicolumn{1}{l}{} & \multicolumn{4}{c}{\textbf{Naive model$^b$}}                                                 \\ \cline{4-7} \cline{9-12} 
Parameter                  & \multicolumn{1}{l}{} & $\theta$             & $\hat{\theta}$       & ESE                  & ASE                  & CR (95\%)            &                      & $\hat{\theta}$       & ESE                  & ASE                  & CR (95\%)            \\ \hline
\multicolumn{2}{l}{\textit{Longitudinal process}} & \multicolumn{1}{l}{} & \multicolumn{1}{l}{} & \multicolumn{1}{l}{} & \multicolumn{1}{l}{} & \multicolumn{1}{l}{} & \multicolumn{1}{l}{} & \multicolumn{1}{l}{} & \multicolumn{1}{l}{} & \multicolumn{1}{l}{} & \multicolumn{1}{l}{} \\ \hline
\textit{Intercept}         & $\beta_0$            & $14.0                $ & $14.01 $               & $0.12 $                & $0.12 $                & $95.4 $                &                      & $13.97$                & $0.12$                 & $0.12                $ & $94.2 $                \\
\textit{Slope}             & $\beta_1$            & $0.17                $ & $0.162$                & $0.074$                & $0.076$                & $95.8$                 &                      &$ 0.198  $              & $0.071 $               & $0.071$                & $93.1   $              \\
\textit{Variability inter} & $\mu_\sigma$         & $0.30                $ & $0.301$                & $0.026$                & $0.027  $              & $95.8 $                &                      & $0.301$                & $0.027$                & $0.026               $ & $94.8 $                \\
\textit{Variability intra} & $\mu_\kappa$         & $-0.23$                & $-0.230 $              & $0.016               $ & $0.017$                & $95.6 $                &                      & $-0.230$               & $0.015$                & $0.016$                & $96.5$                 \\ \hline
\multicolumn{2}{l}{\textit{Transition 0-1}}       & \multicolumn{1}{l}{} & \multicolumn{1}{l}{} & \multicolumn{1}{l}{} & \multicolumn{1}{l}{} & \multicolumn{1}{l}{} & \multicolumn{1}{l}{} & \multicolumn{1}{l}{} & \multicolumn{1}{l}{} & \multicolumn{1}{l}{} & \multicolumn{1}{l}{} \\ \hline
\textit{Current value}     & $\alpha_1^{01}$      & $-0.06$                & $-0.060$               & $0.056$                & $0.054$                & $94.7$                 &                      & $-0.078$               & $0.067$                & $0.057$                & $91.3$                 \\
\textit{Current slope}     & $\alpha_2^{01}$      & $0.0                 $ & $0.006$                & $0.125$                & $0.124$                & $96.0$                 &                      & $-0.093 $              & $1.051$                & $0.147$                & $95.2 $                \\
\textit{Inter variability} & $\alpha_\sigma^{01}$ & $0.50                $ & $0.559 $               & $0.595$                & $0.531$                & $96.6$                 &                      & $0.757 $               & $2.116$                & $0.601$                & $97.0$                 \\
\textit{Intra variability} & $\alpha_\kappa^{01}$ & $0.01                $ & $-0.057$               & $0.780 $               & $0.725$                & $95.2$                 &                      & $-0.176$               & $2.237 $               & $0.763               $ & $94.4 $                \\
\textit{Weibull}           & $\sqrt{\eta^{01}}$   &$ 2.00                $ & $2.022$                & $0.067$                & $0.068$                & $96.2$                 &                      & $1.885$                & $0.172$                & $0.069               $ & $43.3$                 \\
                           & $\zeta^{01}$         & $-4.00               $ & $-4.129 $              & $1.099$                & $1.057$                & $96.2$                 &                      & $-3.919 $              & $2.116$                & $1.168               $ & $93.3$                 \\ \hline
\multicolumn{2}{l}{\textit{Transition 0-2}}       & \multicolumn{1}{l}{} & \multicolumn{1}{l}{} & \multicolumn{1}{l}{} & \multicolumn{1}{l}{} & \multicolumn{1}{l}{} & \multicolumn{1}{l}{} & \multicolumn{1}{l}{} & \multicolumn{1}{l}{} & \multicolumn{1}{l}{} & \multicolumn{1}{l}{} \\ \hline
\textit{Current value}     & $\alpha_1^{02}$      & $-0.10$                & $-0.107$               & $0.067               $ & $0.066$                & $95.2$                 &                      &$ -0.057 $              & $0.041$                &$ 0.043               $ &$ 83.5  $               \\
\textit{Current slope}     & $\alpha_2^{02}$      & $-0.40$                & $-0.440$               & $0.156$                & $0.159$                & $97.2$                 &                      & $-0.269$               & $0.099$                & $0.100$                & $68.6 $                \\
\textit{Inter variability} & $\alpha_\sigma^{02}$ & $0.46                $ &$ 0.421$                & $0.659 $               & $0.648$                & $98.2 $                &                      & $0.392$                & $0.432$                & $0.398$                & $96.1$                 \\
\textit{Intra variability} & $\alpha_\kappa^{02}$ & $0.21                $ & $0.230$                & $0.828$                & $0.843$                & $97.0$                 &                      & $0.149 $               &$ 0.510 $               & $0.526$                & $96.8 $                \\
\textit{Weibull}           & $\sqrt{\eta^{02}}$   & $1.70                $ & $1.709 $               & $0.063$                & $0.071$                & $96.8$                 &                      &$ 1.855$                & $0.048 $               & $0.052$                &$ 10.8  $               \\
                           & $\zeta^{02}$         & $-2.50$                & $-2.454$               & $1.217               $ &$ 1.264$                & $96.6$                 &                      & $-3.237$               & $0.786$                & $0.838               $ & $88.7$                 \\ \hline
\multicolumn{2}{l}{\textit{Transition 1-2}}       & \multicolumn{1}{l}{} & \multicolumn{1}{l}{} & \multicolumn{1}{l}{} & \multicolumn{1}{l}{} & \multicolumn{1}{l}{} & \multicolumn{1}{l}{} & \multicolumn{1}{l}{} & \multicolumn{1}{l}{} & \multicolumn{1}{l}{} & \multicolumn{1}{l}{} \\ \hline
\textit{Current value}     & $\alpha_1^{12}$      & $0.04                $ & $0.049$                & $0.074 $               & $0.072$                & $95.8$                 &                      & $0.011$                & $0.152 $               & $0.088$                & $96.8$                 \\
\textit{Current slope}     & $\alpha_2^{12}$      & $0.02                $ & $0.010$                & $0.170$                & $0.167$                & $96.0$                 &                      & $-0.285$               & $3.514$                & $0.303               $ & $97.8$                 \\
\textit{Inter variability} & $\alpha_\sigma^{12}$ & $-0.12               $ & $-0.065$               & $0.690$                & $0.578 $               & $96.8$                 &                      & $0.582 $               & $6.825$                & $0.847$                & $97.6$                 \\
\textit{Intra variability} & $\alpha_\kappa^{12}$ & $-0.18$                & $-0.207$               & $0.915$                & $0.837$                & $95.8$                 &                      & $-0.565$               & $6.508$                & $1.126$                & $98.3$                 \\
\textit{Weibull}           & $\sqrt{\eta^{12}}$   & $1.70$                 & $1.754$                & $0.139$                & $0.136$                & $94.5$                 &                      & $1.971$                & $0.588$                & $0.150$                & $66.0$                 \\
                           & $\zeta^{12}$         & $-2.20               $ & $-2.622$               & $1.528 $               & $1.399 $               &$ 93.9 $                &                      & $-4.529$               & $8.827$                & $1.865               $ & $86.4$       
                         \\  \hline
\end{tabular}}\\
\footnotesize{
$^a$495 samples with convergence criteria fulfilled; $^b$462 samples with convergence criteria fulfilled.}
\end{table}



\begin{table}[]
\caption{Results of the simulation study for Scenario C, comparing estimates of the joint illness-death model for interval-censored events and the naive joint illness-death model. A total of 500 samples of 1000 subjects were generated with a joint illness-death model with visit times driven by 3C. ASE is the asymptotic standard error, ESE is the empirical standard error and the coverage rate is calculated from the 95\% confidence interval.}
\centering
\label{SC}
\scalebox{0.8}{
\begin{tabular}{lccccccccccc}
\hline
                                                                                                                                                                                                                       
                           & \multicolumn{1}{l}{} & \textbf{}            & \multicolumn{4}{c}{\textbf{Interval censoring$^a$}}                                           & \multicolumn{1}{l}{} & \multicolumn{4}{c}{\textbf{Naive model$^a$}}                                                 \\ \cline{4-7} \cline{9-12} 
Parameter                  & \multicolumn{1}{l}{} & $\theta$             & $\hat{\theta}$       & ESE                  & ASE                  & CR (95\%)            &                      & $\hat{\theta}$       & ESE                  & ASE                  & CR (95\%)            \\ \hline
\multicolumn{2}{l}{\textit{Longitudinal process}} & \multicolumn{1}{l}{} & \multicolumn{1}{l}{} & \multicolumn{1}{l}{} & \multicolumn{1}{l}{} & \multicolumn{1}{l}{} & \multicolumn{1}{l}{} & \multicolumn{1}{l}{} & \multicolumn{1}{l}{} & \multicolumn{1}{l}{} & \multicolumn{1}{l}{} \\ \hline
\textit{Intercept}         & $\beta_0$            & $14.0$                 & $14.00$                & $0.10$                 & $0.09                $ & $93.4 $                &                      & $14.01$                &$ 0.10 $                & $0.10$                 & 93.8                 \\
\textit{Slope}             & $\beta_1$            & $0.17$                 & $0.167$                & $0.062$                & $0.061$                & $94.6$                 &                      & $0.155$                & $0.060$                &$ 0.059$                & $94.2 $                \\
\textit{Variability inter} & $\mu_\sigma$         & $0.30$                 & $0.299 $               &$ 0.021 $               & $0.020$                & $95.6  $               &                      & $0.301 $               & $0.020 $               & $0.020 $               & $95.2$                 \\
\textit{Variability intra} & $\mu_\kappa$         & $-0.23               $ & $-0.230$               & $0.017$                & $0.016$                & $95.2$                 &                      & $-0.231$               & $0.016$                & $0.016$                & $95.0$                 \\ \hline
\multicolumn{2}{l}{\textit{Transition 0-1}}       & \multicolumn{1}{l}{} & \multicolumn{1}{l}{} & \multicolumn{1}{l}{} & \multicolumn{1}{l}{} & \multicolumn{1}{l}{} & \multicolumn{1}{l}{} & \multicolumn{1}{l}{} & \multicolumn{1}{l}{} & \multicolumn{1}{l}{} & \multicolumn{1}{l}{} \\ \hline
\textit{Current value}     & $\alpha_1^{01}$      & $0.20$                 & $0.204$                & $0.046$                & $0.043$                & $93.6$                 &                      & $0.151$                & $0.043$                & $0.042               $ & $77.5 $                \\
\textit{Current slope}     & $\alpha_2^{01}$      & $0.0                 $ & $-0.003$               & $0.078 $               & $0.075$                & $94.4$                 &                      & $-0.047$               & $0.081$                & $0.075               $ & $90.2$                 \\
\textit{Inter variability} & $\alpha_\sigma^{01}$ &$ 0.80                $ & $0.819$                & $0.149$                & $0.148$                & $95.0$                 &                      & $0.583$                & $0.155$                & $0.145               $ & $64.3$                 \\
\textit{Intra variability} & $\alpha_\kappa^{01}$ & $0.01                $ & $0.014$                & $0.203$                & $0.190$                & $93.2 $                &                      & $-0.048 $              & $0.209$                & $0.187               $ & $93.0$                 \\
\textit{Weibull}           & $\sqrt{\eta^{01}}$   & $2.00$                 & $2.006$                & $0.046$                & $0.047$                & $95.0$                 &                      & $1.851$                & $0.045$                & $0.047               $ & $10.4$                 \\
                           & $\zeta^{01}$         & $-7.00               $ & $-7.105$               &$ 0.773   $             & $0.733$                & $94.8$                 &                      & $-5.867$               & $0.722$                & $0.704$                & $64.5$                 \\ \hline
\multicolumn{2}{l}{\textit{Transition 0-2}}       & \multicolumn{1}{l}{} & \multicolumn{1}{l}{} & \multicolumn{1}{l}{} & \multicolumn{1}{l}{} & \multicolumn{1}{l}{} & \multicolumn{1}{l}{} & \multicolumn{1}{l}{} & \multicolumn{1}{l}{} & \multicolumn{1}{l}{} & \multicolumn{1}{l}{} \\ \hline
\textit{Current value}     & $\alpha_1^{02}$      & $0.30                $ & $0.302$                & $0.109$                & $0.097$                & $93.2$                 &                      & $0.305$                & $0.048$                & $0.048$                & $95.8$                 \\
\textit{Current slope}     & $\alpha_2^{02}$      & $0.10                $ & $0.098$                & $0.199 $               & $0.184$                & $92.4$                 &                      & $0.066 $               & $0.079 $               &$ 0.080$                & $92.4 $                \\
\textit{Inter variability} & $\alpha_\sigma^{02}$ & $0.20                $ & $0.129$                & $0.424$                & $0.394$                & $94.2$                 &                      & $0.811$                & $0.152$                & $0.155$                & $1.6$                  \\
\textit{Intra variability} & $\alpha_\kappa^{02}$ & $0.20$                 &$ 0.148$                &$ 0.426$                & $0.394$                & $93.6$                 &                      & $0.141$                & $0.190$                & $0.193               $ & $94.2$                 \\
\textit{Weibull}           & $\sqrt{\eta^{02}}$   & $1.70$                 &$ 1.697 $               & $0.089  $              & $0.086               $ & $93.8$                 &                      & $2.011$                & $0.056$                & $0.052$                &$ 0.0$                  \\
                           & $\zeta^{02}$         & $-8.00$                & $-7.959$               & $1.797$                & $1.608$                & $93.2$                 &                      & $-9.144$               & $0.839$                & $0.836$                & $75.1$                 \\ \hline
\multicolumn{2}{l}{\textit{Transition 1-2}}       & \multicolumn{1}{l}{} & \multicolumn{1}{l}{} & \multicolumn{1}{l}{} & \multicolumn{1}{l}{} & \multicolumn{1}{l}{} & \multicolumn{1}{l}{} & \multicolumn{1}{l}{} & \multicolumn{1}{l}{} & \multicolumn{1}{l}{} & \multicolumn{1}{l}{} \\ \hline
\textit{Current value}     & $\alpha_1^{12}$      & $0.15                $ & $0.158$                & $0.052$                & $0.051$                & $95.2$                 &                      & $0.145$                & $0.038$                & $0.049$                & $99.2$                 \\
\textit{Current slope}     & $\alpha_2^{12}$      & $0.10$                 & $0.101$                & $0.085$                & $0.085               $ & $95.8$                 &                      & $0.025$                & $0.051$                & $0.072               $ & $90.4$                 \\
\textit{Inter variability} & $\alpha_\sigma^{12}$ & $0.80$                 & $0.813$                & $0.192$                & $0.181$                & $94.0$                 &                      & $0.565$                & $0.093$                & $0.140$                & $63.5$                 \\
\textit{Intra variability} & $\alpha_\kappa^{12}$ & $0.10                $ & $0.128$                & $0.227$                & $0.220$                & $94.4 $                &                      & $0.107$                & $0.146$                & $0.198               $ & $99.4$                 \\
\textit{Weibull}           & $\sqrt{\eta^{12}}$   & $1.70                $ & $1.717$                & $0.095$                & $0.092$                & $94.4$                 &                      & $1.725$                & $0.074$                &$0.086$                & $96.2$                 \\
                           & $\zeta^{12}$         & $-4.50$                & $-4.712$               & $1.033$                & $1.007$                & $94.2$                 &                      & $-4.576 $              & $0.718$                & $0.919$                & $99.4 $    
                        \\   \hline
\end{tabular}}\\
\footnotesize{
$^a$498 samples with convergence criteria fulfilled.}
\end{table}

Tables \ref{SA}, \ref{SB} and \ref{SC} displays the results of the simulation studies for each scenario. For each scenario, we compared the results of our model properly handling interval censoring into account with those obtained from the model without taking interval censoring into account. For the three scenarios, the estimation procedure for the model accounting for the interval censoring provided satisfactory results as the bias was minimal, the mean asymptotic and the empirical standard errors were close, and the coverage rates of the 95\% confidence interval were close to the nominal value. When comparing these results with those of the model that does not account for interval censoring, we observe higher biases and underestimation of the standard errors, leading to poor coverage rates. These trends are more pronounced for Scenario B and Scenario C as the interval between visits is larger or the signal increase. Especially for scenario C, we observe large biases for the association parameters between inter-visits variability and event risks. As shown by \cite{leffondre_2013}, the bias is stronger for scenario C because the inter-visits variability is strongly associated with both dementia and death after dementia. Thus, subjects with high variability are more prone to die before the first visit following dementia onset and consequently to be undiagnosed. They are thus considered as died without dementia in the naive analysis.

\section{Application}
The proposed model was then applied to the French prospective 3C Cohort to study the effect of the inter- and intra-visit blood pressure variability on the risk of dementia.

\subsection{The Three-City Cohort}
The 3C study is a population-based prospective cohort which aimed at assessing the relation between vascular factors and dementia in the elderly \citep{3c_2003}. Participants, aged 65 years and older, were randomly selected in 1999 from the electoral lists of three French cities, Bordeaux, Dijon and Montpellier. This analysis was performed on the Bordeaux subsample because the follow-up was longer. This Bordeaux sample included 2104 participants at baseline who were followed every 2-3 years up to 20 years. At each visit, systolic blood pressure (SBP) was measured two or three times. Each measurement was taken in a seated position after a rest period.\\
The final sample of analysis included 1788 non-demented participants at baseline for whom there is at least one SBP measurement and sex, educational level and APOE are known. Among these participants 493 developed dementia, 1002 died without previous dementia diagnosis and 401 died after dementia diagnosis. Participants were 74 years old at baseline on average, 61\% were women, 62\% had an educational level lower than secondary school and 19\% carried the APOE allele.

\subsection{Specification of the model}
The aim of the study was to evaluate the impact of within and between visits blood pressure variability on the risk of transition to dementia and death (before and after dementia). We estimated the proposed joint model defined by \eqref{MixedApp} using the age as time scale. The mean trajectory of blood pressure was described over time by a linear location-scale mixed effect model. The individual time trend of the marker was modelled by a linear trend. The baseline hazard functions of each transition were defined by a Weibull function. 
The models for transition intensities for each event depended on both inter and intra-visit variabilities, the individual current value and the current slope of the blood pressure, and they were also adjusted for sex (male versus female), educational level (higher than secondary school ($\geq 10$ years of study or not) and on carrying the APOE allele:
\begin{equation}
\left\{
    \begin{array}{ll}
         Y_{ijl} =  \widetilde{Y}_i(t_{ij}) + \epsilon_{ij} + \nu_{ijl} = \beta_0 + b_{0i} +(\beta_1+b_{1i})\times t_{ij}+\epsilon_{ij} + \nu_{ijl}, \\
        \epsilon_{ij} \sim \mathcal{N}(0,\sigma_i^2) \hspace{3mm} \text{with} \hspace{3mm} \log(\sigma_i)  = \mu_\sigma + \tau_{\sigma i},\\
        \nu_{ijl} \sim \mathcal{N}(0,\kappa_i^2) \hspace{3mm} \text{with} \hspace{3mm} \log(\kappa_i)  = \mu_\kappa + \tau_{\kappa i},\\
        \lambda_{i}^{kl}(t)=\lambda_{0}^{kl}(t) \exp
        \left(\gamma_1^{kl}Sex_i + \gamma_2^{kl}Edu_i + \gamma_3^{kl}Apoe4_i + \alpha_{1}^{kl}\tilde{y}_i(t)+\alpha_{2}^{kl}\tilde{y'}_i(t)+
     \alpha_{\sigma }^{kl} \sigma_i +  \alpha_{\kappa }^{kl} \kappa_i \right)
    \end{array}
\right.  
 \label{MixedApp}
\end{equation}
with $t = \frac{age-65}{10}$ and the random effects $b_i$ and $\tau_i = (\tau_{\sigma i}, \tau_{\kappa i})^\top$ are supposed to be independent. The estimation was performed with $S1 = 1000$ and $S2 = 5000$ draws of QMC to ensure a greater accuracy.\\

\subsection{Results}

\begin{table}[]
\centering
\caption{Parameter estimates of the joint model on the 3C-Bordeaux data ($N=1788$).}
\label{App}
\begin{tabular}{llccc}
\hline
\multicolumn{2}{l}{\textbf{Parameter}}                               & \textbf{Estimate} & \textbf{Standard error} & \textbf{p-value} \\ \hline
\multicolumn{2}{l}{\textit{Dementia (transition 0-1)}}               &                   &                         &                  \\
              & BP current value                                     & $-0.071$           & 0.051                   & 0.163           \\
              & BP slope                                             &$ -0.070$           & 0.077                   & 0.364           \\
              & BPV between-visits                                   & 0.597              & 0.256                   & 0.019            \\
              & BPV within-visit                                     & $-0.084$            & 0.335                   & 0.802            \\
              & Woman                                                & $-0.027$            & 0.109                  & 0.804            \\
              & Education ($\geq 10$ years)                                & $-0.230$            & 0.098                  & 0.020            \\
              & APOE4                                                & 0.404             & 0.110                   & $<$0.001           \\
\multicolumn{2}{l}{\textit{Death without dementia (transition 0-2)}} & \textit{}         & \textit{}               & \textit{}        \\
              & BP current value                                     & $-0.086$            & 0.072                   & 0.235            \\
              & BP slope                                             & $-0.196$            & 0.103                   & 0.058            \\
              & BPV between-visits                                   & 0.396             & 0.331                   & 0.232            \\
              & BPV within-visit                                     & 0.319             & 0.411                   & 0.437            \\
              & Woman                                                & $-0.760$            & 0.1097                  & $<$0.001           \\
              & Education ($\geq10$ years)                         & $-0.087 $           & 0.104                   & 0.404           \\
              & APOE4                                                & $-0.015$            & 0.138                   & 0.915            \\
\multicolumn{2}{l}{\textit{Death after dementia (transition 1-2)}}   &                   &                         &                  \\
              & BP current value                                     & 0.035            & 0.048                   & 0.464            \\
              & BP slope                                             & 0.009             & 0.079                   & 0.909            \\
              & BPV between-visits                                   & $-0.231$             & 0.218                  & 0.290            \\
              & BPV within-visit                                     & $-0.154$            & 0.298                   & 0.607           \\
              & Woman                                                & $-0.556$            & 0.101                  & $<$0.001           \\
              & Education ($\geq 10$ years)                 & 0.081             & 0.096                   & 0.404            \\
              & APOE4                                                & $-0.067$            & 0.109                   & 0.538            \\
\multicolumn{2}{l}{\textit{Longitudinal submodel}}                   &                   &                         &                  \\
              & Intercept                                            & 13.90             & 0.081                   & $<$0.001           \\
              & Time                                                 & 0.211             & 0.053                   & $<$0.001           \\
              & Intercept of between-visits variability              & 0.299             & 0.015                   & $<$0.001           \\
              & Intercept of within-visits variability               & $-0.228$          & 0.010                   & $<$0.001           \\ \cline{1-5} 
\end{tabular}
\end{table}

    \begin{align}
    \Sigma & =  \left[\begin{array}{cccc}
Var(b_{0i}) &  &  &   \\
Cov(b_{0i},b_{1i}) & Var(b_{1i}) &  &  \\
Cov(b_{0i},\tau_{\sigma i}) & Cov(b_{1i},\tau_{\sigma i})  & Var(\tau_{\sigma i}) & \\
Cov(b_{0i},\tau_{\kappa i}) & Cov(b_{1i},\tau_{\kappa i}) & Cov(\tau_{\sigma i},\tau_{\kappa i}) & Var(\tau_{\kappa i}) 
\end{array}\right]\\ 
& = \left[\begin{array}{ccccc}
4.57_{(0.50)} &  &  & &  \\
-1.86_{(0.61)} & 1.22_{(0.86)} &  &  &\\
0 &  0 & 0.07_{(0.01)} &  \\
0 &  0 & 0.01_{(6e-3)} & 0.07_{(6e-3)} 
\end{array}\right]
\label{DM}
\end{align}

Table~\ref{App} provides estimates of the regression parameters from the joint model and equation (\ref{DM}) presents the covariance matrix of the random effects and their standard errors computed through the Delta-Method. Mean blood pressure increased with time ($\beta_1 = 0.211$, $p <0.001$, table~\ref{App}). Both between and within visits variances of the residual part were heterogenerous between the subjects ($\widehat{Var}(\tau_{\sigma i}) = 0.07$, $sd = 0.01$ and $\widehat{Var}(\tau_{\kappa i}) = 0.07$, $sd = 0.006$) which confirms the importance of taking heteroscedasticity into account (Figures S1 on Supplementary Materials). The risk of dementia was lower for subjects with a high educational level (Hazard Ratio: $HR = 0.79$, $IC = [0.66;0.96]$) but 50\% higher for subjects carrying the APOE4 allele ($HR = 1.50$, $IC = [1.21;1.86]$). There was no effect of the sex ($p = 0.80$). Adjusting for these covariates, the risk of dementia increased with the between-visits blood pressure variability ($HR = 1.82$,  $IC = [1.10;3.00]$) but we found no significant effect of within-visit blood pressure variability, the current slope and the current value. Then, considering both the risk of death after dementia and without dementia,  the risk of death was lower for women ($HR = 0.47$, $IC = [0.38;0.58]$ without dementia and $HR = 0.57$, $IC = [0.47;0.70]$ with dementia) but none of the other covariates and none of the characteristics of the blood pressure trajectory was significantly associated with the risk of death.

\subsection{Goodness-of-fit}

\begin{figure}
        \includegraphics[scale=0.55]{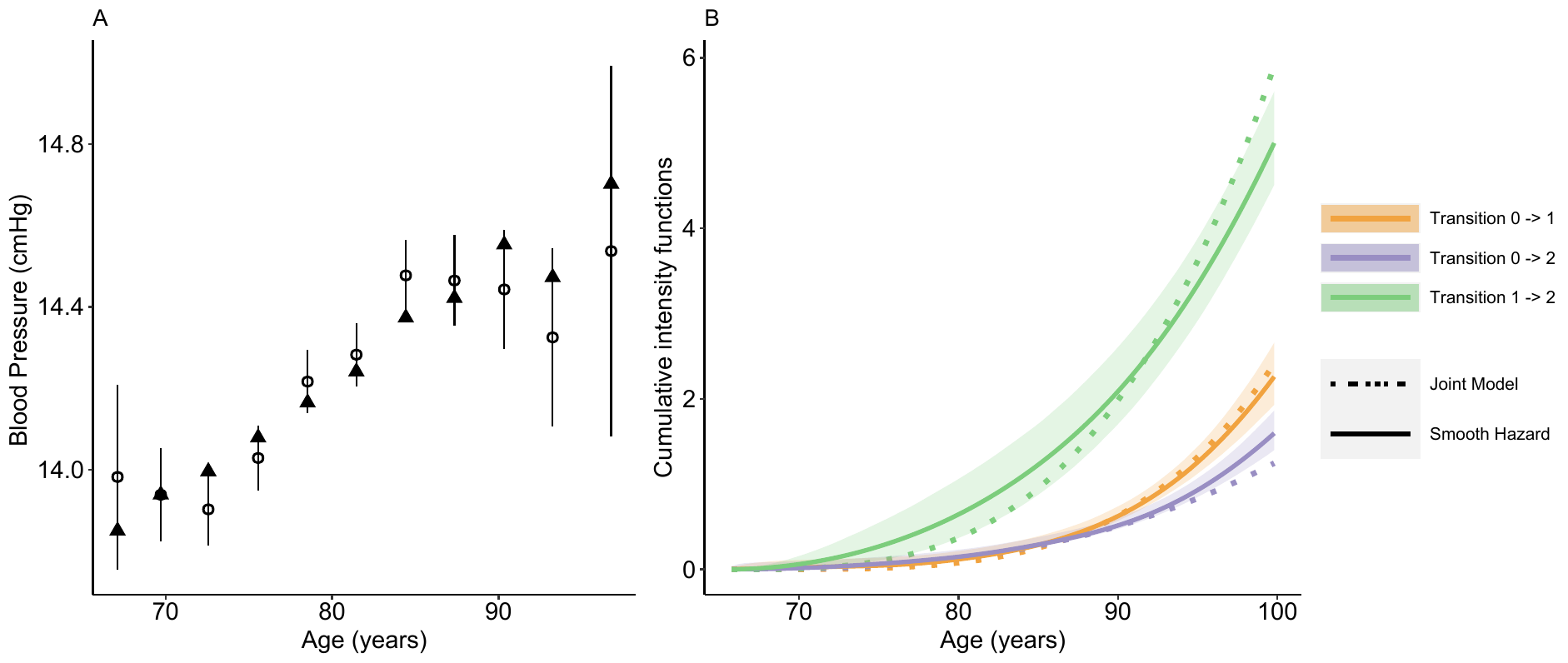}
        \caption{Goodness-of-fit of the model for the 3C-Bordeaux cohort ($N=1788$): A. Mixed effects submodel: Comparison between predicted value of the marker from the joint model (black triangles) and the observations (mean in white circles with 95\% confidence interval (by 3-year age intervals). B. Illness-death model: Comparison between the predicted cumulative hazard function from the joint model for each transition and a non-parametric estimator with 95\% confidence interval (Illness-death model estimated by penalized likelihood accounting for interval censoring with the R package \texttt{SmoothHazard}).}
        \label{Gof}
\end{figure}

To assess fit of the longitudinal sub-model we computed the empirical Bayes estimates of the random effects and the predicted value of the marker for each individual at their respective visiting times. Figure \ref{Gof} A compares the mean of marker predictions at each visit time to the mean of the observed measurements. It shows that the joint model adequately fit the trajectory of the blood pressure.\\

\noindent Then to evaluate the fit of the illness-death submodel, we plugged the empirical Bayes estimates of the random effects in the formula for the risk functions to compute the predicted cumulative hazard function in a grid of times. Figure \ref{Gof} B compares the mean of this predicted cumulative hazard function for each transition to a non-parametric estimator obtained using the SmoothHazard package \citep{touraine_2017}. It highlights that the proposed joint model adequately fitted each transition.

\section{Discussion}
We proposed a new joint model for semi-competing interval censored events, and longitudinal data with heteroscedasticity assuming two subject-specific residual variances: one within-visit and one between-visits. This model allows to study the impact of the residual variabilities on the risks of the events. Simulations studies demonstrated the good performance of the estimation procedure dealing with interval censoring and emphasized biased estimates obtained with a naive model estimated by imputing the middle of the censoring interval for the time of dementia onset and assuming that all the subjects who died without dementia diagnosis were really not demented at death. The R-package \texttt{LSJM} has been developped to allow the estimation of such models and is available on Github at the following link: \url{https://github.com/LeonieCourcoul/LSJM}.\\
The analysis of the 3C cohort with this model has shown that a high between-visits blood pressure variability increases the risk of dementia but the within-visit variability seems to have no impact. This suggests that intra-visit variability could mainly quantifies measurement error rather than true variability. This does not support the hypothesis that intra-visit variability, which is easily measured could be an interesting indicator of the dementia risk.\\
This model relies on some hypothesis. Regarding missing data for marker measurements, it is important to note that this model deals with the informative dropouts due to dementia and death thanks to the joint modeling of the risks of death and dementia. But it assumes that missing marker measurements for other causes are missing at random. Moreover, this model relies on a Markovian assumption but semi-Markovian hypothesis could be considered. However, \cite{rouanet_2016} shown previously that mortality among subjects with dementia depends more on age than on duration of dementia, suggesting that the Markovian hypothesis is better.\\
Joint models that account for inter-visits and intra-visit subject-specific residual variance are highly valuable for examining the relationship between both variabilities of markers or risk factors and the risk of health events across various areas of medical research. Furthermore, the proposed model can be extended to consider not only inter-visits and intra-visit variability but also long-term versus short-term variability via a temporal window with different measurements times for the short-term variability. It could be useful to study the variability of a marker during different period of time, for instance the short-term variability could be the intra-day variability and the long-term variability the variability during all the follow-up.\\

\section*{Acknowledgements}

This work was funded by the French National Research Agency (grant ANR-21-CE36 for the project "Joint Models for Epidemiology and Clinical research").\\
This PhD program is supported within the framework of the PIA3 (Investment for the Future). Project reference: 17-EURE-0019.\\
Computer time was provided by the computing facilities MCIA (Mésocentre de Calcul Intensif Aquitain) at the University of Bordeaux and the University of Pau and Pays de l’Adour.\\
We thank Christophe Tzourio for providing access to the data from the 3C Study, for discussions and motivation of this work. The 3C Study was funded by Sanofi-Synthélabo, the FRM, the CNAM-TS, DGS, Conseils Régionaux of Aquitaine, Languedoc-Roussillon, and Bourgogne; Foundation of France; Ministry of Research- INSERM “Cohorts and biological data collections" program; MGEN; Longevity Institute; General Council of the Côte d’Or; ANR PNRA 2006 (grant ANR/ DEDD/ PNRA/ PROJ/ 200206–01-01) and Longvie 2007 (grant LVIE-003-01); Alzheimer Plan Foundation (FCS grant 2009-2012); and Roche. The Three-City Study data are available upon request at \url{e3c.coordinatingcenter@gmail.com}.

\bibliographystyle{unsrtnat}
\bibliography{arxiv.bib}
\newpage
\include{Surparxiv}

\end{document}

%% file: Surparxiv.tex
\begin{center}
\LARGE{Supplementary Material \\}
\LARGE{Joint model for interval-censored semi-competing events and longitudinal data with subject-specific within and between visits variabilities}
\vspace{3mm}
\normalsize
Léonie Courcoul$^{1*}$, Catherine Helmer$^1$,\\
Antoine Barbieri$^1$,  and Hélène Jacqmin-Gadda$^1$

\noindent$^1$Univ. Bordeaux, INSERM, Bordeaux Population Health, U1219, France\\
\vspace{2mm}
\end{center}
\section*{Appendix A: Process of data generation}
For each individual, the age of dementia onset and death were generated in the following steps:

\begin{enumerate}
    \item Generate $T_{01i}$ and $T_{02i}$ using the Brent’s univariate root-finding method according to the following proportional hazards models:
\begin{equation}
    \lambda_{i}^{kl}(t)=\lambda_{0}^{kl}(t) \exp \left(\alpha_{1}^{kl}\tilde{y}_i(t)+
     \alpha_{2}^{kl}\tilde{y}'_i(t)+ \alpha_{\sigma }^{kl} \sigma_i +  \alpha_{\kappa }^{kl} \kappa_i \right)
    \label{SimuSurv}
\end{equation}
with $\lambda_{0}^{kl}(t) = \eta^{kl} t^{\eta^{kl}-1}e^{\alpha_{0}^{kl}}$ being a Weibull function.
    \item If $T_{01i} > A_{0i} + C_i$ and $T_{02i} > A_{0i} + C_i$ (with $C_i$ the right censoring and $A_{0i}$ the age at inclusion) then the individual is free of any event at the end of the follow-up: $T_i = A_{0i} + C_i$, $\delta_i^{dem} = 0$, $\delta_i^{death} = 0$ and $L_i =  max(t_{ij})$ (the last visit).
    \item Else, if $T_{02i} < T_{01i}$ and  $T_{02i} \le A_{0i} + C_i$, then the individual is dead at $T_{02}$ without dementia diagnosis: $T_i = T_{02}$, $\delta_i^{dem} = 0$, $\delta_i^{death} = 1$ and $L_i =   max(t_{ij})$.
    \item Else, if $T_{01i} \le T_{02i}$ and $T_{01i} \le A_{0i} + C_i$, then the individual developed dementia and we generate the time to death from dementia $T_{12i}$ according to the intensity model for transition 1-2; 
    \begin{enumerate}
        \item if $T_{12i} > A_{0i} + C_i $ then the subject is alive at the end of the follow-up:
            \begin{itemize}
                \item if $T_{01i} >  max(t_{ij})$: the dementia was not diagnosed so $T_i = A_{0i} + C_i $, $\delta_i^{dem} = 0$, $\delta_i^{death} = 0$ and $L_i = max(t_{ij})$.
                \item else, dementia is diagnosed: $T_i = A_{0i} + C_i $, $\delta_i^{dem} = 1$, $\delta_i^{death} = 0$, $L_i = max\{t_{ij} /t_{ij} \le T_{01i}\}$ (ie. the last visit before $T_{01i})$ and $R_i = min\{t_{ij} / t_{ij} \ge T_{01i})$ (ie. the first visit after $T_{01i}$).
            \end{itemize}
            \item else ($T_{12i} \le A_{0i} + C_i $), then the subject died at $T_{12i}$:
            \begin{itemize}
                \item if $T_{01i} > max(t_{ij})$ : the dementia was not diagnosed so $T_i = T_{12}$, $\delta_i^{dem} = 0$, $\delta_i^{death} = 1$ and $L_i = max(t_{ij})$.
                \item if $T_{01i}$ and $T_{12i}$ are in the same interval between two consecutive visits, then the subject was not diagnosed: $T_i = T_{12i}$, $\delta_i^{dem} = 0$, $\delta_i^{death} = 1$ and $L_i =  max\{t_{ij} / t_{ij} < T_{12i}\}$ (ie the last visit before death).
                \item else the subject is diagnosed: $T_i = T_{12i}$, $\delta_i^{dem} = 1$, $\delta_i^{death} = 1$, $L_i = max\{t_{ij} / t_{ij} \le T_{01i}\}$ and $R_i = min\{t_{ij} / t_{ij} \ge T_{01i}\}$.
            \end{itemize}

        \end{enumerate}
        \item Delete the measurements of the longitudinal marker measured after $T_{02i}$ or $T_{12i}$.
    \end{enumerate}
    \newpage
\section*{Appendix B: Histograms of inter and intra-visit variabilities}
\begin{figure}[!h]
        \includegraphics[scale=0.8]{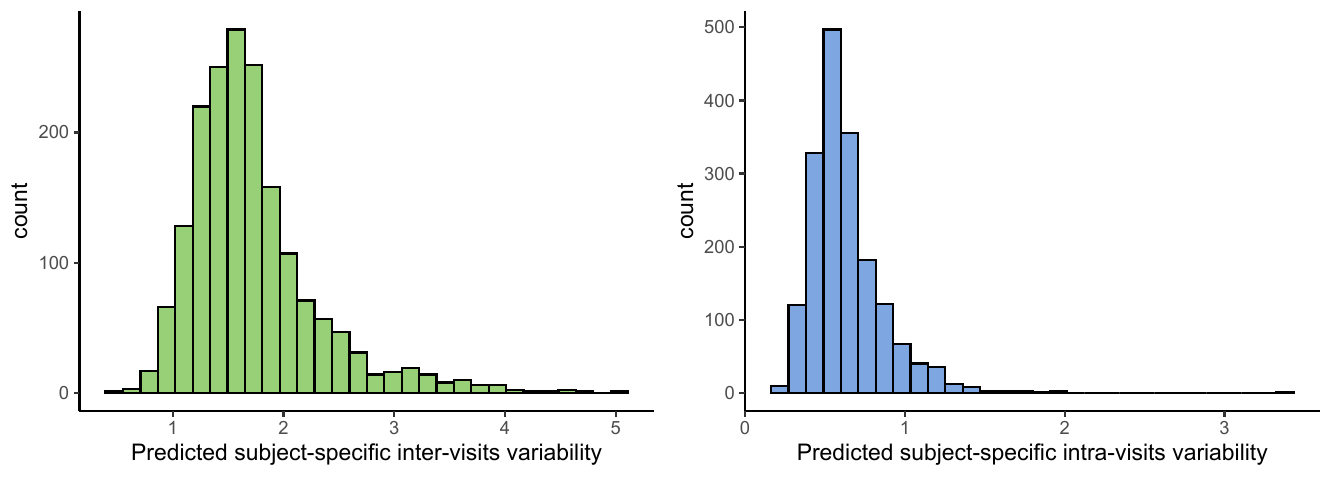}
        \caption{Histogram of predicted subject-specific inter-visits variability (left) and intra-visits variability (right) from the 3C-Bordeaux cohort ($N=1788$).}
\end{figure}